\documentclass{article}[11pt]
\usepackage{graphicx}
\usepackage{placeins}
\usepackage{subfig}
\usepackage{amsmath}

\newcommand{\miktex}{\hbox{Mik\kern-.15em\TeX}}

\newcommand{\rd}{{\rm d}}

\newcommand{\oh}{{\frac{1}{2}}}

\newcommand{\alp}{\alpha}
\newcommand{\etal}{{\em et al.}}

\newcommand{\nn}{\nonumber}
\newcommand{\APNY}[1]{Ann. Phys. (N.Y.) {\bf {#1}}}

\newcommand{\NPA}[1]{Nucl. Phys. {\bf A{#1}}}
\newcommand{\PLB}[1]{Phys. Lett. {\bf B{#1}}}

\newcommand{\PRC}[1]{Phys. Rev. C {\bf {#1}}}

\providecommand{\rjpurl}[1]{\bgroup\footnotesize\ttfamily #1\egroup}
\begin{document}
\input{epsf.sty}

\title{LOW ENERGY $\alp-\alp$ SEMIMICROSCOPIC POTENTIALS}

\author{M. Lassaut$^{1,2}$, F. Carstoiu$^2$, and  V. Balanica$^2$ \\
$^1$  Institut de Physique Nucl\'eaire\\
  IN2P3-CNRS, Universit\'e Paris-Sud 11\\
  F-91406 Orsay Cedex, France\\
\vspace{1mm}
$^2$ National Institute for Nuclear Physics and Engineering,\\
P.O.Box MG-6, RO-077125 Bucharest-Magurele, Romania}

\date{}
\maketitle
\begin{center}
(Received \today )
\end{center}

\begin{abstract}
The $\alp-\alp$ interaction potential is  obtained within the  double folding model  with
 density-dependent Gogny effective interactions as input. The one nucleon knock-on
exchange kernel including recoil effects is localized using the Perey-Saxon
prescription at zero energy. The Pauli forbidden states
 are removed  thanks to successive supersymmetric transformations.
Low energy experimental phase shifts, calculated from the variable phase approach,
  as well as  the energy and width
of the first $0^+$ resonance in $^8$Be are reproduced with high accuracy.
\end{abstract}
\bigskip


%\pacs{01.30.-y, 01.30.Ww, 01.30.Xx, 99.00.Bogus}

\hyphenation{rjp-ar-ti-cle}

%%%%%%%%%%%%%%%%%%%%%%%%%%%%%%%%%%%%%%%%%%%%%%%%%%%%%%%%%%%%%%%%%%%%%%%%%%%%%%%
%Please, do not remove nor uncomment the following lines!
%%%%%%%%%%%%%%%%%%%%%%%%%%%%%%%%%%%%%%%%%%%%%%%%%%%%%%%%%%%%%%%%%%%%%%%%%%%%%%%
%\RJPVolume{57}{2012}
%\RJPNumber{9-10}
%\RJPPages{}{}
%\colontitle{}
%\date{}
%\dedication{}
%\domaintitle{}
%%%%%%%%%%%%%%%%%%%%%%%%%%%%%%%%%%%%%%%%%%%%%%%%%%%%%%%%%%%%%%%%%%%%%%%%%%%%%%%

{\section{INTRODUCTION \label{int} }}

In last time there is an increasing interest in understanding the properties of
 $\alpha$-matter  mainly due to the believe that this type of hadronic matter
occurs in astrophysical environment in unconfined form.
In the debris of a supernova explosion, a substantial fraction of hot and dense matter
resides in $\alpha$-particles and therefore the equation of state of subnuclear matter
is essential in simulating the supernova collapse and explosions and is also
important for the formation of the supernova neutrino signal \cite{lattimer}.

 The basic ingredient in the calculation of the ground state alpha matter  \cite{miscar09}
as well in the $\alp$-cluster model of nuclei \cite{sofianos1}
is the $\alp-\alp$ interaction potential. This has been studied extensively using
both local and nonlocal interactions.
Among the most important are those using the resonating group model (RGM) \cite{spuy, wilderm},
the energy and angular momentum independent
potential model of Buck, Friedrich and Wheatley \cite{buck} and the
phenomenological potential of Ali and Bodmer \cite{AB66}. There have been proposed several
versions of the Ali-Bodmer potential: a Gaussian potential with a stronger repulsive
component by Langanke and M\"uller \cite{langanke}, as well a version with a
softer repulsive component by Yamada and Schuck \cite{schuck}. All these models
predict potentials quite different in
strength and range but all are claimed to reproduce experimental
data up the the breakup threshold.

Microscopic RGM calculations by Schmid and Wildermuth \cite{wilderm} lead to the
important conclusion that due to the compact structure and the large binding
energy the radius of the $\alp$-particle stays essentially the same during the
compound system formation and therefore the polarization effects could be
neglected. This observation substantiates the idea of calculation of a
$\alp-\alp$ potential from the double folding model.

 We propose in this paper to generate the $\alp-\alp$ potential within the double
folding model using the Gogny force as
input. Previously  Sofianos \etal \cite{sofianos2} derived the $\alp-\alp$ potential
using the energy density formalism based on Skyrme effective interaction.

However, the potential issued from double-folding calculation, even
corrected by knock-on exchange terms,  is generally too deep due to the presence of forbidden
 bound states. These states have a clear interpretation within the RGM model: they are
redundant solutions giving fully
antisymmetrized wave functions that vanish identically.
  These latter bound states  are eliminated thanks to successive  supersymmetric
 transformations   as given in  \cite{baye1},
  which preserves the continuous spectrum (phase-shift) and resonances \cite{Baye0}.

 In section {\bf 2} we present the derivation of the   $\alpha-\alpha$ interaction. In section {\bf 3} the derivation
  and the properties of supersymmetric partner are presented.  Our conclusions are given in section {\bf 4}.

\vspace{1cm}

\section{Bare $\alpha-\alpha$ interaction : double-folding with Gogny forces}

Since the potentials providing saturation at lower densities of the alpha matter are highly schematic
 (infinite repulsive short-range interactions) we turn to a calculation of the
bare $\alpha-\alpha$ interaction based on the double-folding method for two ions
at energies around the barrier, starting with realistic densities of the
$\alp$-particle and modern effective nucleon-nucleon interactions.

Within the double-folding model \cite{carst92} the interaction between two
alpha clusters is calculated as a convolution of a local two-body potential
$v_{nn}$ and the single particle densities of the two clusters,  namely
\begin{equation}
v_{\alp\alp}(\vec{r})=\int d\vec{r}_1 \int d\vec{r}_2
\rho_{\alp}({r}_1)\rho_{\alp}({r}_2)v_{nn}(\rho,\vec{r}-\vec{r}_1+\vec{r}_2)
\label{dfold}
\end{equation}

The effective  $n-n$ interaction $v_{nn}$ is taken to be density-dependent
as expected from a realistic interaction. It depends on the density $\rho$
 of the nuclear matter where the two interacting nucleons
are embedded. For the sake of simplicity, we choose Gaussians interactions
 in order to have the most  tractable analytical calculations.
    A candidate satisfying this requirement is provided by the Gogny forces \cite{Gogny73}.
In this paper, we will  report results using  three main parametrizations of the Gogny
interaction \cite{Gogny73}, denoted D1 and D1S \cite{DG80}  as well as the most recent variant, labeled D1N \cite{CGS08}.

We  remind  that the standard form  of  the Gogny interactions is,
\begin{equation}\begin{aligned}
v_{nn}(r)&=\sum_{i=1}^2(W+BP_{\sigma}-HP_{\tau}-MP_{\sigma}P_{\tau})e^{-r^2/\mu_i^2}
\\
&+t_3(1+x_0P_{\sigma})\rho^{\gamma} \left(\frac{\vec r_1+\vec r_2}{2}\right)\delta(\vec r_1-\vec r_2)
\label{gognyvnn}
\end{aligned}
\end{equation}
where   $\vec{r}=\vec{r_1}-\vec{r_2}$, and the coefficients
  $W,B,H,M$   refer to  the usual notations for the spin/isospin mixtures and
$P_{\sigma,\tau}$ are the spin/isospin exchange operators. The spin-orbit
component,  present in the original formulation,  is ignored here
as it is not material   for the $\alp-\alp$ system.

 For the sake of consistency, i.e. working with Gaussian interactions, we consider
  Gaussian  one-body density for  the $\alp$-particle
\begin{equation}
\rho_\alp(r)=4\left ( \frac{1}{\pi b^2}\right )^{3/2} e^{-r^2/b^2} \ .
\label{rhoalf}
\end{equation}
 In Eq.(\ref{rhoalf}) the oscillator parameter $b$ is adjusted on
the root mean square radius of the $\alpha$-particle (r.m.s.) given by $<r^2>^{1/2}=b \sqrt{3/2}$ which
has to be compared  to the value 1.58 $\pm$0.002 fm, extracted from a
Glauber analysis of experimental interaction cross sections \cite{AlK96}.

A more involved density matrix was derived by  Bohigas and Stringari \cite{bohigas} who included
short range correlations  starting from a Jastrow wave function and evaluated
the one-body density matrix by using
the perturbation expansion of \cite{gaudin} at a low order. The diagonal
component of the density matrix so
far obtained is not far from our density (\ref{rhoalf}), and since we
want to keep the results as simply as
possible we use Eq. (\ref{rhoalf}). We have checked that the density Eq.
(\ref{rhoalf}) reproduces the experimental charge form factor \cite{frosch} up to
$q^2\sim 2 fm ^{-2}$ momentum transfer.

Antisymmetrization of the density dependent term in the Gogny force is obtained
at follows. Consider the operator,
\begin{equation}
\cal{O}=\it{\left(1+x_0 P^\sigma\right)(1-P^\sigma P^\tau P^x)}\\
\end{equation}
Since $\delta$ acts only in S-states, one can take safely $P^x=1$ and using the usual
algebra of the exchange operators one obtains,
\begin{equation} v_d^\rho(r_{12})=t_3 \left(1+\frac{x_0}{2} \right) \rho^\gamma \delta(r_{12}),
\end{equation}
and,
\begin{equation}
v_{ex}^\rho(r_{12})=-\frac{t_3}{4}(1+2x_0)\rho^\gamma\delta(r_{12}) \ .
\end{equation}
 The interest is that  the total contribution from the density dependence, is calculated
from
\begin{equation}
v^{\rho}(r_{12})=\frac{3}{4}t_3\rho^\gamma\delta(r_{12})
\end{equation}
and is independent of the value of the spin mixture $x_0$.
Therefore we take $x_0=1$.
 The direct spin-isospin independent effective $n-n$ force in the Gogny parametrization
\cite{miscar09} reads:
\begin{equation}
v_{00}^{\rm d}(\vec{r}_1-\vec{r}_2)=\oh\sum_{i=1}^2(4W_i+2 B_i-2
H_i-M_i)e^{-|\vec{r}_1-\vec{r}_2|^2/\mu_i^2}
+\frac{3}{2}t_3\rho^{\gamma}
%\left [\rho\left(\oh(\vec{r}_{1}+\vec{r}_{2})\right)\right ]^{\gamma}
\delta(\vec{r}_1-\vec{r}_2)
\end{equation}

Inserting the Gaussian density distribution (\ref{rhoalf}) in
the double folding integral (\ref{dfold}) and using a generalization of the Campi-Sprung
prescription  \cite{campi72}  for the overlap density similar to
the one proposed in \cite{DLMV} for $\alpha$-nucleus scattering
\begin{equation}
\rho(1,2)=\left (\rho_\alp(\vec{r}_1-\frac{1}{2}\vec{s})\rho_\alp(\vec{r}_2+\frac{1}{2}
\vec{s})\right )^{\frac{1}{2}},
\end{equation}
where $\vec{s}=\vec{r}_1+\vec{r}-\vec{r}_2$ is the $n-n$ separation in the
heavy-ion coordinate system \cite{carst92}. With this approximation, the overlap
density does not exceeds the density of the normal nuclear matter at complete
overlap and goes to zero when one of the interacting nucleon is far from the
other. We obtain the local $\alp-\alp$ potential,
\begin{equation}\begin{aligned}
v_{\alp\alp}(r)&=4\sum_{i=1}^2(4W_i+2 B_i-2H_i-M_i)\left
(\frac{\mu_i^2}{\mu_i^2+2b^2}\right )^{3/2}e^{-{r^2}/{(\mu_i^2+2b^2)}}\\
&+\frac{3}{2}t_3\frac{4^{\gamma+2}}{(\gamma+2)^{3/2}(\sqrt{\pi}b)^{3(\gamma+1)}}
e^{-\frac{\gamma+2}{4b^2}{r^2}}
\label{vaa}
\end{aligned}
\end{equation}
which includes both direct and exchange arising from the density dependent component of the force.

The derivation of the knock-on exchange component corresponding to the finite
range component of the effective interaction is more involved.
 It is convenient to start  from the DWBA matrix element of the
exchange operator :
\begin{equation}
\hat{U}_{ex}\chi=\sum_{\alpha \beta}<\phi_\alpha(\vec{r}_1)\phi_\beta(\vec{r}_2)|v_{ex}(s)P_{12}^x|
\phi_\alpha(\vec{r}_1)\phi_\beta(\vec{r}_2)\chi(\vec{R})>
\end{equation}
where the sum runs over the single-particle wave functions of occupied states
in the projectile (target) and $\chi(\vec{R})$ is the wave function for relative motion.
After some  algebra (see details in \cite{carslass96}),
we arrive at,
\begin{equation}
\hat{U}_{ex}\chi=\int U_{ex}(\vec{R},\vec{R}^\prime)\chi(\vec{R}^\prime)d\vec{R}^\prime\nn
\end{equation}
where the kernel $U_{ex}(\vec{R},\vec{R}^\prime)$ is given by,
\begin{flalign}
U_{ex}(\vec{R},\vec{R}^\prime)=U_{ex}(\vec{R}^+,\vec{R}^-)=\mu^3v_{ex}(\mu R^-)
\int \rho_1(\vec{X}+\delta_1\mu\vec{R}^-,\vec{X}-\delta_1\mu\vec{R}^-)&\nn \\
\times \rho_2(\vec{X}-\vec{R}^+-\delta_2\mu\vec{R}^-,\vec{X}-\vec{R}^++\delta_2\mu\vec{R}^-)
d\vec{X}&
\label{eq211}
\end{flalign}
where $\vec{R}^+=(\vec{R}+\vec{R}^\prime)/2,~~\vec{R}^-=\vec{R}-\vec{R}^\prime$ and
$\rho(\vec{r},\vec{r}^\prime)$ is the one-body matrix  density. The $\delta_i=1-\frac{1}{A_i}$ accounts for recoil effects. The equation (\ref{eq211}) already tells us that the
range of non-locality $\vec{R}^-$ is  $\sim \mu^{-1}$ .
In the case of the $\alp-\alp$ interaction we have
\begin{equation}\label{nonlocal}\begin{aligned}
U_{\alp\alp}^{\rm ex}(\vec{R},\vec{R}^\prime)&=8v_{00}^{\rm ex}(2R^-)
\int \rho_\alp(\vec{X}+\frac{3}{2}\vec{R}^-,\vec{X}-
\frac{3}{2}\vec{R}^-)\\
&\times  \rho_\alp(\vec{X}-\vec{R}^+-\frac{3}{2}\vec{R}^-,\vec{X}-\vec{R}^++\frac{3}{2}\vec{R}^-) d\vec{X}
\end{aligned}\end{equation}

The local equivalent potential is well approximated \cite{Peierls80} by the
lowest order term of the Perey-Saxon approximation. For high energy
and a heavy target the $\alp$-nucleus potential reads,
\begin{equation}\label{localiz}\begin{aligned}
U_L(R)&=\int e^{i\vec{K}\vec{R}^-}U_{\alp\alp}^{\rm ex}(\vec{R}+\frac{1}{2}\vec{R}^-,\vec{R}^-)d\vec{R}^- \\
&=4\pi\int \rho_\alp(X)\rho_\alp(|\vec{R}-\vec{X}|)d\vec{X}\\
&\times \int v_{00}^{\rm ex}(s)\hat{j}_1(\hat{k}_1(X)\frac{3}{4}s)\cdot\hat{j}_1(\hat{k}_2(|\vec{R}-\vec{X}|)\frac{3}{4}s)\\
&\times j_0(K(R)s/2)s^2ds
\end{aligned}\end{equation}
where $K(R)$ is the usual WKB local momentum for the relative motion,
\begin{equation}
K^2(R)=\frac{2\mu}{\hbar^2}(E_{c.m.}-U_D(R)-U_L(R))
\end{equation}
and $U_D$ is the direct term including the
nuclear and Coulomb potentials. Truly speaking, the classical momentum is
defined only for energies where $K^2(R)\geq0$. At
under-barrier energies, $K(R)$ is imaginary in the region $R_1<R<R_2$, where $R_{1,2}$ are the
classical turning points of the total potential, and the Bessel function $j_0$ above should be
replaced by $j_0(ix)=\sinh(|x|)/|x|$. In Eq. (\ref{localiz}) the function
$\hat{j}_1(x)=3j_0(x)/x$ arises from the Slater approximation of the mixed
density.

%We sketch below the derivation of the exchange part of the
%potential. The exchange part of the Gogny potential reads
%\begin{equation}
%v_{00}^{\rm ex}(\vec{r}_1-\vec{r}_2)=\frac{1}{4}\sum_{i=1}^2(W_i+2 B_i-2
%H_i-4M_i)e^{-|\vec{r}_1-\vec{r}_2|^2/\mu_i^2}
%-\frac{3}{4}t_3\rho^{\gamma}
%\left [\rho\left(\oh(\vec{r}_{1}+\vec{r}_{2})\right)\right ]^{\gamma}
%\delta(\vec{r}_1-\vec{r}_2)
%\end{equation}
\begin{figure}[!ht]
\centering
\includegraphics[width=0.75\textwidth]{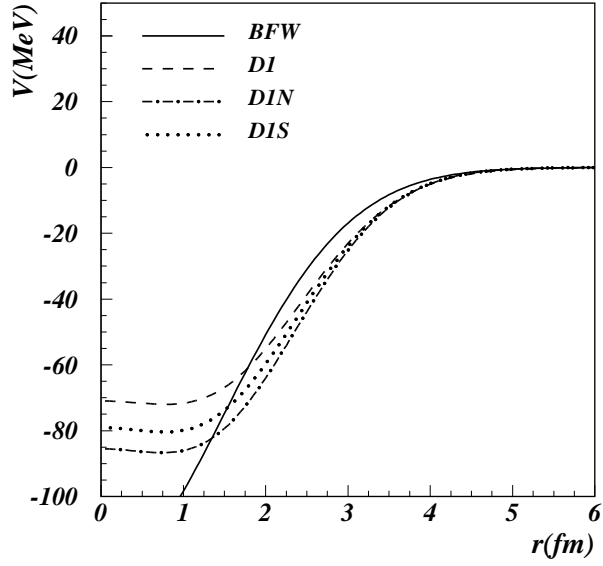}
\caption{Folding  $\alp-\alp$ potentials (including knock-on exchange) obtained from three
parametrizations of the Gogny effective interaction. The Coulomb component is
omitted. The phenomenological BFW potential is plotted for comparison.}
\label{figpot1}
\end{figure}

In the particular case of the $\alp-\alp$ system the one body density matrix can
be evaluated exactly from $0S$ HO orbitals,
%The matrix-density corresponding to the density (\ref{rhoalf}) is
\begin{equation}
\rho_\alp(\vec{r},\vec{r}^\prime)=4\left ( \frac{1}{\pi b^2}\right )^{3/2} e^{-(r_+^2+\frac{1}{4}r_-^2)/b^2}
\end{equation}
where
\begin{equation}
\vec{r}_+=\oh(\vec{r}+\vec{r}^\prime),~~~ \vec{r}_-=\vec{r}-\vec{r}^\prime
\end{equation}
Explicitly we have,
\begin{equation}\label{eq3b}\begin{aligned}
\rho_\alp(\vec{X}+\frac{3}{2}\vec{R}_-,\vec{X}-\frac{3}{2}\vec{R}_-)&=4\left ( \frac{1}{\pi b^2}\right )^{3/2}
 e^{-(\vec{X}+\frac{9}{4}\vec{R}_-^2)/b^2}\\
\rho_\alp(\vec{X}-\vec{R}_+-\frac{3}{2}\vec{R}_-,\vec{X}-\vec{R}_+
+\frac{3}{2}\vec{R}_-) &=4\left ( \frac{1}{\pi b^2}\right )^{3/2}
e^{-\left [ (\vec{X}-\vec{R}_+^2)+\frac{9}{4}\vec{R}_-^2\right ]/b^2}
\end{aligned}\end{equation}
Using the convolution techniques we obtain the compact expression
of the non-local kernel,
\begin{equation}
U_{\alp\alp}^{\rm ex}(\vec{R},\vec{R}^\prime)=
-4\left ( \frac{2}{\pi b^2}\right )^{3/2}\sum_i^2(W_i+2 B_i-2
H_i-4M_i)e^{-\oh \left ( \frac{8}{\mu_i^2}+\frac{9}{b^2}\right )R_-^2}
e^{-\frac{1}{2b^2}R_+^2}
\end{equation}
Adopting the short-hand notation
\begin{equation}
\frac{1}{\beta_i^2}= \frac{8}{\mu_i^2}+\frac{9+\frac{1}{4}}{b^2}
\end{equation}
and using the integral identity
\begin{equation}
\int d\vec{s}e^{-\alp^2s^2}e^{i\beta\vec{s}\cdot\vec{K}}=
\left ( \frac{\pi}{\alp^2}\right )^{3/2}e^{-(\beta K/2\alp)^2}
\end{equation}
the local equivalent of the nonlocal kernel  in the lowest order of
the Perey-Saxon procedure is obtained as, \cite{misicu},

% \begin{eqnarray}
%v_{\alp\alp}^{\rm ex}(r)&=&-32\sum_i(W_i+2 B_i-2H_i-4M_i)
%\left (\frac{\beta_i}{b}\right )^3 e^{-\frac{1}{2b^2}
%\left[1-\frac{1}{4}\left (\frac{\beta_i}{b}\right )^2\right ]r^2}\nn\\
%&\times &e^{-\oh K^2\beta_i^2}e^{\frac{1}{2} i\left (\frac{\beta_i}{b}\right )^2\vec{K}\cdot\vec{r}}
%\end{eqnarray}
\begin{equation}\label{eq2b}\begin{aligned}
v_{\alp\alp}^{\rm ex}(r)=&-32\sum_i(W_i+2 B_i-2H_i-4M_i)
\left (\frac{\beta_i}{b}\right )^3 e^{-\frac{1}{2b^2}
\left[1-\frac{1}{4}\left (\frac{\beta_i}{b}\right )^2\right ]r^2} \\
&\times e^{\pm\oh |K|^2\beta_i^2}
\left\lbrace
\begin{array}{ccc}
e^{-\oh \left (\frac{\beta_i}{b}\right )^2|{K}|{r}}
& {\rm for} & K^2 < 0\\
\cos\left [ \oh\left (\frac{\beta_i}{b}\right )^2|{K}|{r}\right ]
& {\rm for} & K^2 \geq 0
\end{array}
\right.
\end{aligned}\end{equation}
Thus we have a sub-barrier branch ($K^2<0$) and an over-barrier one ($K^2>0$)
for the real part of the local exchange potential. The potentials depicted in Fig.
(\ref{figpot1}) are obtained by applying the localization procedure at
$E_{c.m.}=0$.
The  deep potential of Buck \etal (BFW) \cite{buck} which has has two $\ell=0$ bound states
located at -72.79 MeV and -25.88 MeV respectively is displayed for comparison. We notice that
all Gogny forces give very close potentials  at the surface.

\begin{figure}[!ht]
  \centering
  \subfloat[]{\label{figpot2}\includegraphics[width=0.5\textwidth]{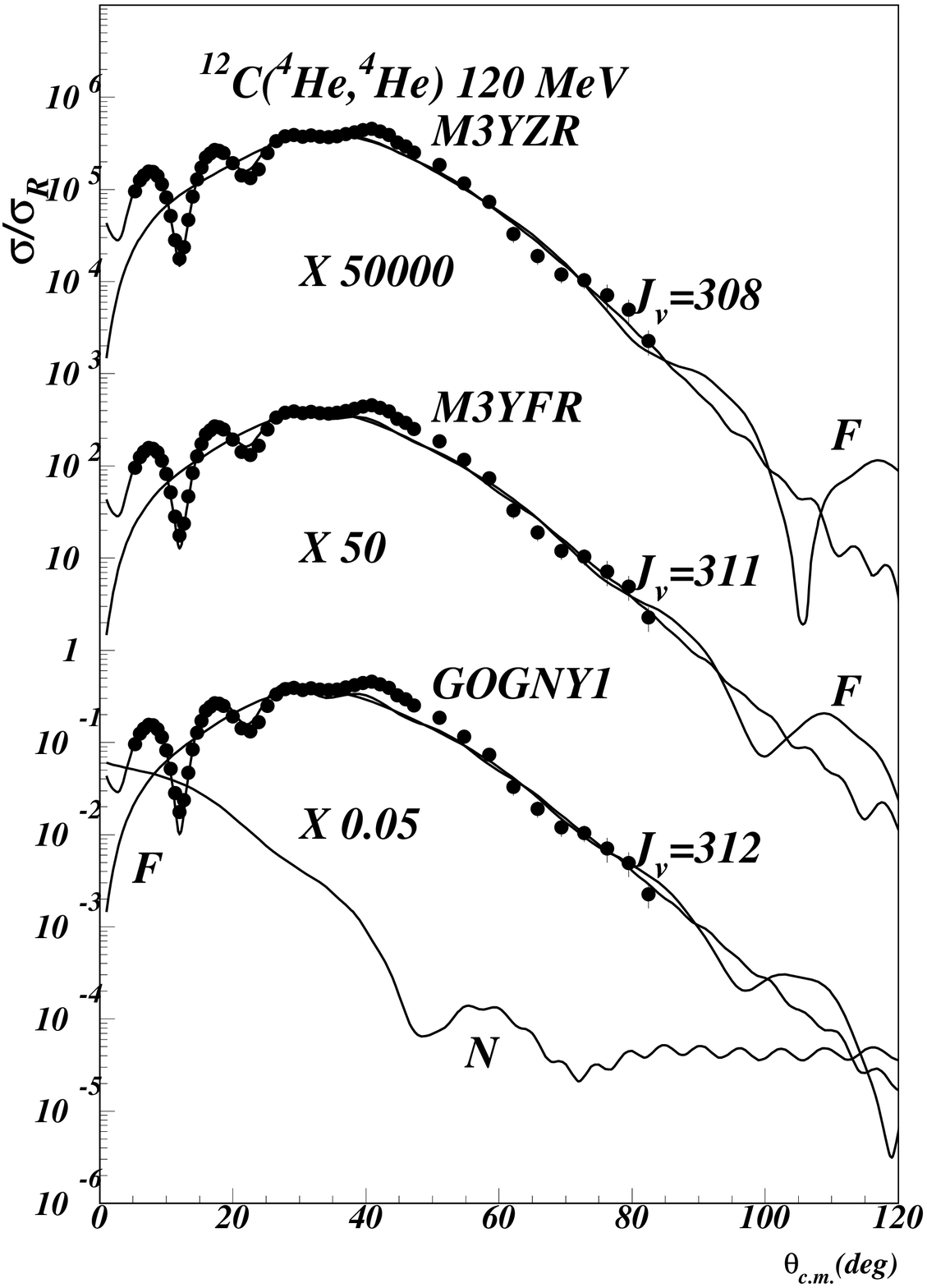}}
 \subfloat[]{\label{figpot3}\includegraphics[width=0.5\textwidth]{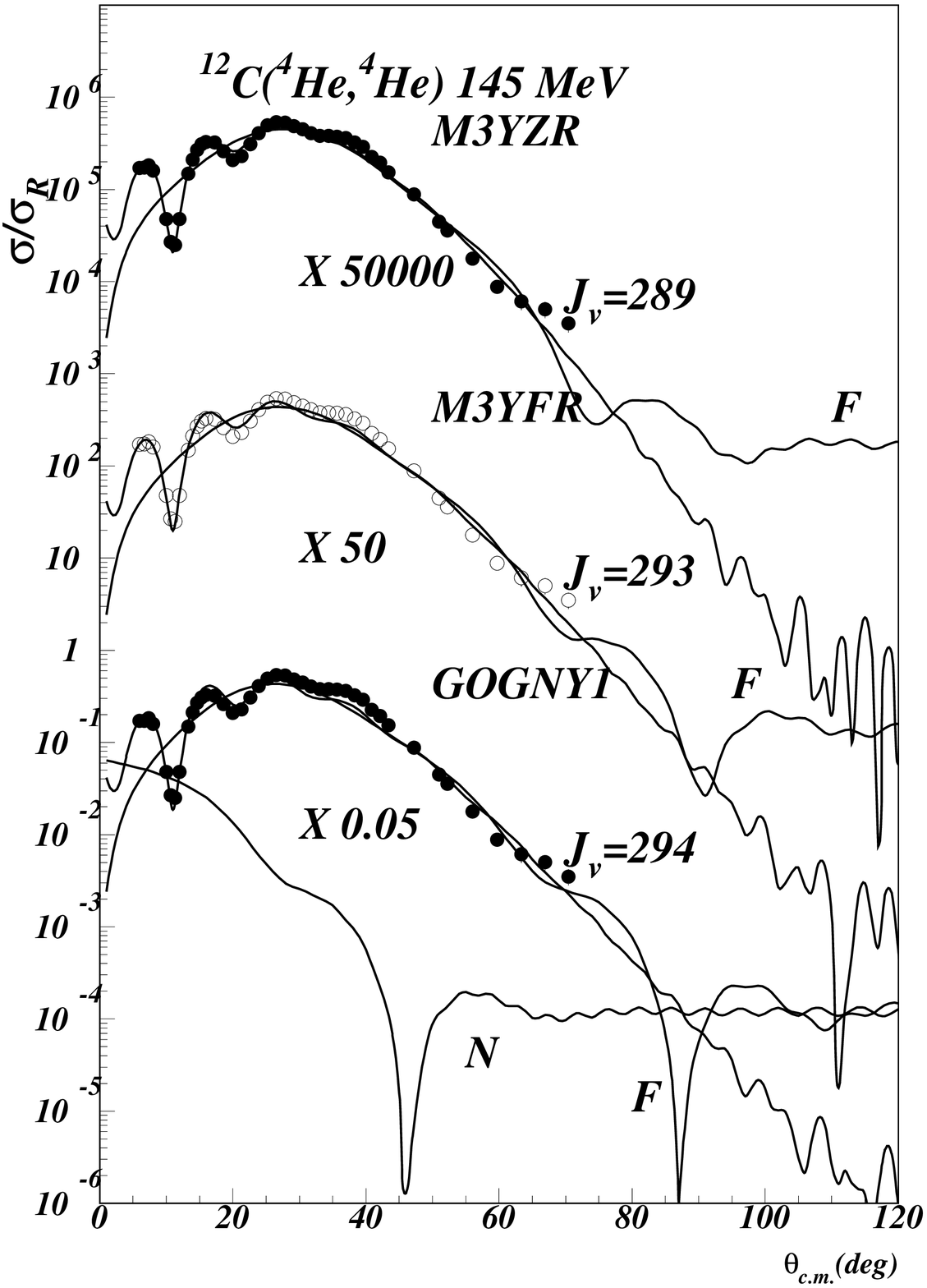}}
 %  \caption{}
% \label{figpo}
\caption{ Test of the heavy ion potential calculated with the D1
parametrization of the Gogny effective
interaction on the high energy $\alpha$ scattering. The results
are comparable with those obtained with the
 zero range and finite range versions of the well known M3Y interaction.
\label{figpo} }
\end{figure}

The potentials are tested against high energy experimental data in Figure \ref{figpo}. The
results with   the
Gogny force $D1$, are labeled Gogny1 on the figure. Curves labeled F/N are
the far side/near side components of the scattering amplitude. The real and imaginary
form factors calculated with
Eq. (\ref{localiz}) are slightly renormalized to match the experimental data.

\section{Supersymmetric partners of the bare interactions}

%ccccccccccccccccc

Once with have obtained the bare interactions by folding including the local
equivalent of the knock-on exchange kernel we notice that the resultant deep
 potential has two non-physical bound states.
 Also, there are several candidates reproducing qualitatively well the experimental data
 (see Figure \ref{figpo}). Therefore, the question of the uniqueness of the potential is raised.
 The question of forbidden states is well-known and has been studied in the
 supersymmetry approach in \cite{Baye0}.
These states should be removed in order to obtain a physically meaningful
$\alp-\alp$ potential.

 In this section we describe  the method used to remove two bound states
using the formalism of Baye \cite{baye1,baye2} and of Baye and Sparenberg \cite{bayeS3} see also refs.
\cite{baye4,baye5}.
We give the straightforward generalization  of   equations (3.3) and (3.5) of \cite{baye2} to the case
where two bound states are removed simultaneously.
%This has been made in  \cite{bayeS3}
%which gives the expression of the final potential .
Our potential is expected to be energy-dependent because of the Perey-Saxon approximation.
Generally this latter energy dependence is linear and we should apply the derivation
 of Sparenberg, Baye and Leeb \cite{SBL} for linearly energy-dependent potentials.
 For the sake of simplicity we take the Perey-Saxon at zero energy and consider
  the standard derivation of supersymmetric partners \cite{bayeS3}.

Here we consider the case in absence of Coulomb potential. In fact, we will see
 further, the results are not, in a certain measure,
  affected by the presence of the Coulomb potential.

\subsection{Notations}
We consider the Schr\"odinger equation for the $\ell$-wave

\begin{equation}
\left(\frac{\rd^2}{\rd r^2} +
    \frac{2 \mu}{\hbar^2} ( E-V(r) ) -\frac{\ell (\ell+1)}{r^2} \right)
     \psi_{\ell}(E,r) =  0
\label{Schr}
\end{equation}
where $ \psi_{\ell}(E,r)$ is called the regular solution which is uniquely defined, as usual \cite{newt,CS}, by
the Cauchy condition $\lim_{r \to 0} \psi_{\ell}(E,r) r^{-\ell -1}=1$. It behaves for positive values of $E$ as
$\psi_{\ell} \propto \sin(k r - \ell \pi/2 + \delta_{\ell}(k))$
when  $r\to\infty$ ($k=\sqrt{2 \mu E/\hbar^2}$), provided that $V(r)$ satisfies the
integrability condition \cite{CS}
\begin{equation}
 \int_b^{+\infty} \vert V(r) \vert \rd r  <  \infty, \quad b>0, \qquad
            \int_0^{\infty} r  \vert V(r) \vert \rd r  <  \infty
\label{int1}
\end{equation}
Here, the $ \delta_{\ell}(k)$'s are the phase shifts.
In all equations $\mu$ denotes the reduced mass of the system and $E$ the c.m. energy.
When the potential possesses bound states  labeled $E_0<E_1< \ldots <E_N \leq 0$
(the number of which is finite when the potential satisfies  the integrability  condition Eq.(\ref{int1})) we can define their normalization $C_j$ (relative to $E_j$) constant as
\begin{equation}
\frac{1}{C_j} = \int_0^{\infty} dr \ \psi_{\ell}(E_j,r)^2  \ .
\label{norm}
\end{equation}

 Note that the integrability condition (\ref{int1})  discards the Coulomb potential.
 In fact, we will see further, the results are not, in a certain measure, affected by
 the presence of the Coulomb potential.
%  but this not affect in principle
% our discussion.

 It  is worth to recall that the exact phase $\delta_{\ell}$ can
be calculated by using the variable phase
method of Calogero \cite{Calo}.
With this method, the phase-shift is obtained
by solving a first order differential equation
\begin{equation}
\frac{\partial}{\partial r} \delta_{\ell}(k,r) =-\frac{v(r)}{k}
\ (u_{\ell}(k r)  \cos(\delta_{\ell}(k,r))+  w_{\ell}(k r)
\sin(\delta_{\ell}(k,r)) )^2 \ ,
\label{cal}
\end{equation}
with $\delta_{\ell}(k,0) = 0$ as boundary condition. In equation (\ref{cal})
$v(r)=2 \mu V(r)/\hbar^2$ is the reduced potential.
  The phase-shift
is  given by the limit $\delta_{\ell}(k) =\lim_{r \to \infty} \delta_{\ell}(k,r)$.

The regular $u_{\ell}(k r)$ and irregular $w_{\ell}(k r)$
solutions of  Eq.(\ref{Schr}) for $v \equiv 0$ are denoted, respectively,
\begin{equation}\begin{aligned}
u_{\ell}(x) & =  \sqrt{\frac{\pi x}{2}} J_{\ell+1/2}(x) \nonumber\\
w_{\ell}(x) & =  -  \sqrt{\frac{\pi x}{2}} Y_{\ell+1/2}(x) \nonumber
\end{aligned}\end{equation}
in terms of the    Bessel    functions $J_{\nu},Y_{\nu}$ of order $\nu$,
given in \cite{erd}.
We have $u_{\ell}(x)=x j_{\ell}(x)$ where $j_{\ell}$ is the spherical Bessel function of order $\ell$.
 %In the present work, use is made of the variable phase method to determine the exact
%phases.
 For $\ell=0$ we have  $u_0(x)=\sin(x) $ and $w_0(x)=\cos(x)$.
%In presence of Coulomb potential,
% we subtract to the phase, Eq.(\ref{cal}), generated by the nuclear potential plus screened Coulomb,
%the phase generated by the pure Coulomb potential.

Note that for potentials in the class (\ref{int1}) the Levinson theorem,
  ( see  \cite{newt,CS} and its extension to singular potentials in \cite{Swan} )
applies. We have, except for a bound state at zero energy, $\delta_{\ell}(k=0)-\delta_{\ell}(k=\infty)= n_{\ell} \pi$
where $\delta_{\ell}$ is the exact phase (\ref{cal}) and $n_{\ell}$  denotes the number of bound states,
in the $\ell$-wave.

%ccccccccccccccccc

% As these potentials are "regular" i.e. integrable ($\in L^1(R)$  and having a rapid decrease,
% the Levinson theorem applies.  We found a phase of $2 \pi$ at zero energy instead
%of zero expected in the absence of bound state.
%We have to remove these forbidden extra bound state and use
%the well-know supersymmetry.

\subsection{Phase-equivalent potentials}

 In this subsection we remind  the method used to remove two bound states
using the formalism of Baye \cite{baye1,baye2} and of Baye and Sparenberg \cite{bayeS3}. We follow closely the derivation given in refs. \cite{baye4,baye5}.
%All formulas are given in refs. \cite{baye4,baye5} and we report the results here for the case
%of two bound states.

Starting  with the bare potential $v(r)=(2 \mu V(r)/\hbar^2)$ then the phase equivalent
 potential $v^{(1)}(r)$, with the ground state removed is given by,
\begin{equation}
v^{(1)}(r)=v(r)-2 \frac{\rd^2}{\rd r^2} \ln \int_0^r \rd t \ \psi_{\ell}(E_0,t)^2 \
\label{pot1}
\end{equation}
and the corresponding regular solution for $v^{(1)}$ is,
\begin{equation}
\psi_{\ell}^{(1)}(E,r)= \psi_{\ell}(E,r) -  \psi_{\ell}(E_0,r) \frac{\int_0^r \rd t \ \psi_{\ell}(E,t) \  \psi_{\ell}(E_0,t)}{ \int_0^r \rd t \ \psi_{\ell}(E_0,t)^2 }
\label{psi1}
\end{equation}
%we use the expressions depicted in \cite{baye4,baye5} for the remove of the two non-physical bound states.

The potential $v^{(1)}(r)$ behaves near $r=0$ like $2 (2 \ell+3)/r^2$. This is
due to its definition Eq.(\ref{pot1}) taking into account that
$\psi_{\ell}(E_0,r) \simeq r^{\ell+1}$ at the vicinity of zero.

Removing the next bound state at $E_1$ we have,
\begin{equation}
v^{(2)}(r)=v(r)-2 \frac{\rd^2}{\rd r^2} \ln det(M(r))
\label{pot2t}
\end{equation}
where   $M$ is the $2 \times 2 $ matrix

\begin{equation}
M=\left[\begin{matrix} L_{E_0,E_0}(\ell,r) & L_{E_0,E_1}(\ell,r) \cr
\ L_{E_1,E_0}(\ell,r) & L_{E_1,E_1}(\ell,r)
\end{matrix}\right]
\end{equation}
with
\begin{equation}
L_{E_i,E_j}(\ell,r)=L_{E_j,E_i}(\ell,r)=\int_0^r \rd t \ \psi_{\ell}(E_i,t) \ \psi_{\ell}(E_j,t) \ .
\end{equation}

Clearly the determinant of the matrix $M$ behaves like $r^{4 \ell+10}$ at the vicinity of zero
and the resulting potential has a singularity
$(8 \ell+20)/r^2$ at the vicinity of zero.

On the other hand, the regular solution can be written in a compact form \cite{baye4,baye5}

\begin{equation}
 \psi_{\ell}^{(2)}(E,r)=\frac{det(\tilde M(r))}{det(M(r)}
\label{psiS}
\end{equation}
where we have defined

\begin{equation}
\tilde M=\left[\begin{matrix} \psi_{\ell}(E,r) & L_{E,E_0}(\ell,r) &  L_{E,E_1}(\ell,r) \cr
 \psi_{\ell}(E_0,r) &  L_{E_0,E_0}(\ell,r) &  L_{E_0,E_1}(\ell,r) \cr
 \psi_{\ell}(E_1,r) & \ L_{E_1,E_0}(\ell,r) & L_{E_1,E_1}(\ell,r)
\end{matrix}\right]
\end{equation}
% insert here figure figpot10

\begin{figure}[!ht]
\centering
\includegraphics[width=0.8\textwidth]{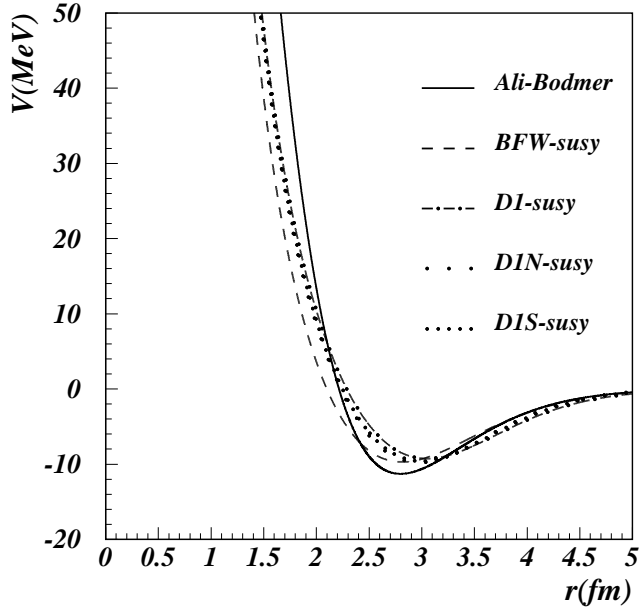}
\caption{The supersymmetric partners of the renormalized bare BFW and Gogny interactions
are compared with Ali-Bodmer phenomenological interaction. We have checked that
original phase shifts and the 0$^+$ resonance properties are conserved}
\label{figpot10}
\end{figure}

\subsection{Uniqueness of the potential}

The present discussion is made discarding  the Coulomb potential.
But we expect that our conclusions remain true as well.
The experimental $\alpha-\alpha$ phase-shift $\ell=0$ are known at discrete
energies up to the breakup threshold \cite {afzal}. Also the properties of the
first $0^+$ resonance in $^8$Be have been measured by Benn \etal \cite{benn}.
%We dispose of the experimental $\alpha-\alpha$ phase-shift $\ell=0$ at different energies.
%Also we know that there exist a resonance at $0.0092$ MeV the width of which is $6$ eV.
  If the experimental S-wave phase-shifts satisfy the condition
$\delta_{\rm exp}(k=0)-\delta_{\rm exp}(k=\infty)=0$, where $k^2=2 \mu E_{cm}/\hbar^2$,
  then  the underlying potential, satisfying  the integrability condition (\ref{int1}),
   has no bound state. It is a  consequence of the Levinson theorem (see above).
Consequently, the  potential is uniquely determined from the phase-shift $\delta_{\ell=0}(k)$,
 given for all positive energies \cite{newt,CS}. The resonance should be at the right place without any fit.

In practical cases, a serious source of uncertainty comes from the fact that
the  phase shifts are  known at a limited number of discrete energies.
Also,  the bare potentials  constructed in the above section
are too  deep and have  two non-physical bound S-states.
Such deep potentials are not unique: indeed their reconstruction from  Gelfand-Levitan or Marchenko  procedure
\cite{newt,CS} includes the S-wave phase-shifts at all positive energies, the bound states and
the corresponding normalization constants Eq.(\ref{norm}).

We dispose of four free parameters namely, the bound state energies $E_1,E_2$ and the associated normalization
 constants $C_1,C_2$. We have to adjust them on the position and width of the resonance
 which eliminates two free parameters.
 The potential is not unique and we have a two-parameters family of solutions.
The supersymmetric transformation described above implies a singularity at
the origin which is that of a centrifugal barrier
of angular momentum $L=2 N$, $N$ corresponding to the number of removed bound states, here
$L=4$ as two bound states are removed.

In a recent paper \cite{ghost} it was advocated that the supersymmetric transformation
 increases the angular momentum by a factor of two in the sense that the Jost
  function $F_{\ell}(k)$,
  %(Eq.\ref{jost}),
  of the starting potential, becomes after removing the
  bound state $E_j=k_j^2 \hbar^2/(2 \mu)$,
\begin{equation}
\tilde F_{\ell+2}(k) =\frac{k^2}{k^2+k_j^2} \ F_{\ell}(k)
\end{equation}

This latter study was made in absence of Coulomb potential.
This implies that the $S_{\ell}$ matrix of the primitive potential is exactly the $S_{\ell+2}$
matrix of the SUSY partner (the potential obtained by removing one bound state)
We then expect that the Calogero phase of the SUSY partner, calculated for the $\ell$-wave
is $-2 \pi/2=-\pi$. For two bound state we will have $-4 (\pi/2)=-2 \pi$.

We stress  the fact that the S-wave Calogero equation (\ref{cal}),  used
to calculate
 the phase shift for a potential having a singularity at the origin
starts from  a modified  boundary condition.
Let be $  \nu (\nu+1)/r^2 \ , \nu \ne -1/2$ the behavior of the singular potential at small distances,
 the Cauchy condition $\delta_{\ell=0}(k,r) =0 $ at small $r$ is changed.
  This comes from the fact that the Calogero variable phase $\delta_{0}(k,r)$ is defined by
$$\delta_0(k,r)=- k r + \arctan \left( \frac{\psi_{0}(k,r)}{ \psi'_{0}(k,r)} \right)  \simeq
- k r + \arctan \left( \frac{k r}{\nu+1} \right)  \simeq - k r \frac{\nu}{\nu+1} $$
so that we start from $\delta_0(k,r) =  -k r \nu/(\nu+1) $.

We have calculated the difference of phase between our deep potentials and the supersymmetric partners
 when two bound states are removed and found $-2 \pi$, even in the presence of the Coulomb potential.
 To conclude our deep potentials supposed to reproduce the experimental phase
have all the same S matrix. This latter is preserved by the supersymmetric
transformations (and then the resonance) and  the resulting SUSY partners 
have the same S matrix but
for the angular momentum L = 4. However, when all bound states of the deep
potential are
removed thanks to supersymmetry the resulting potential is expected to be
unique in
the following sense. If the deep potential supports N bound states of fixed angular momentum $\ell$,
then
the supersymmetric partner, obtained by setting all normalization
constants $C_j$ , j = 1, 2, ...N to infinity \cite{ghost}, is unique, depending only
on the number
N of bound states, which determine the singularity at the origin of the
SUSY partner.

\subsection{Numerical details}

%{\red{\bf ALL IS NEW mais bacl\'e et je dois revoir }}

Consider the physical potentials discussed in section {\bf 2}.
These potentials reproduce reasonably well the experimental phase-shift but fail to reproduce the properties of the
first $0^+$ resonance in $^8$Be. This is true also for the Ali-Bodmer and BFW potentials. We correct this deficiency
by adjusting a global multiplicative factor $\lambda$ and judge the success of our model if $\lambda\approx 1$.

We first calculate the S-wave phase shift $\delta_0(k)$ for the effective potential
\begin{equation}
V_{\rm eff}(r) =V(r)+ 4 e^2  \ \frac{erf( 3 r/4)}{r}
\end{equation}
with $e^2 =  1.43998 $ MeV fm. The screened Coulomb potential arises from the
finite size charge distributions in
the $\alp$-particle. We calculate also the phase $\delta_{0C}(k)$ for the pure
Coulomb potential $V_c(r)=4e^2/r$ and
assume that the difference $\tilde \delta_0(k)=\delta_0(k)-\delta_{0,C}(k)$ is
the nuclear phase shift in the presence of Coulomb potential.

We integrate Eq.(\ref{cal}) up to 500 fm in steps $h=0.001$ fm and
reproduce the exact value of the phase
$\delta_{0C}$ ($\delta_{0C}^{exact}=arg \Gamma(1+i\eta)$) with high precision. The optimum value of the
parameter $\lambda$ is obtained from a grid search around
unity with a continuous refinement of the grid  step  $h_{\lambda}=10^{-3}-10^{-5}$ and keep the value for
which $\sin^2\tilde \delta_0(k)=1$, near the required energy of
0.092 MeV. Note that varying the third decimal of $\lambda$  varies the position of the resonance
 by $5.10^{-4}$. We found values of $\lambda$ close to unity (see Figs.(4) and (5)).
% put here figpot4.eps
\begin{figure}[!ht]
\centering
\includegraphics[width=0.7\textwidth]{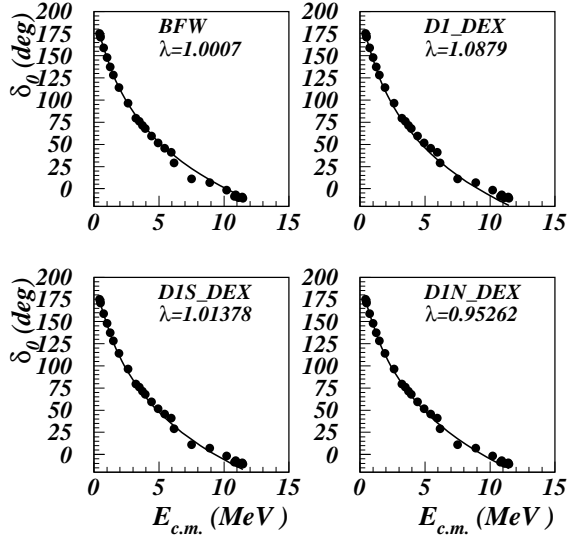}
\label{figpot4}
\caption{The S-state phase shift calculated with bare folding potentials
including direct and exchange components (DEX) are
compared with the BFW results. The parameter $\lambda$ indicate the
renormalization constant.}
\end{figure}

Using henceforth the renormalized potential by the multiplicative factor $\lambda$,
the bound state wave functions for
the redundant 0S and 1S states are calculated using a high precision Numerov scheme. The SUSY potentials are
then calculated using Eq.(\ref{pot2t}) and shown already in Fig (3).

%put here figpot5.eps
\begin{figure}[!ht]
\centering
\includegraphics[width=0.65\textwidth]{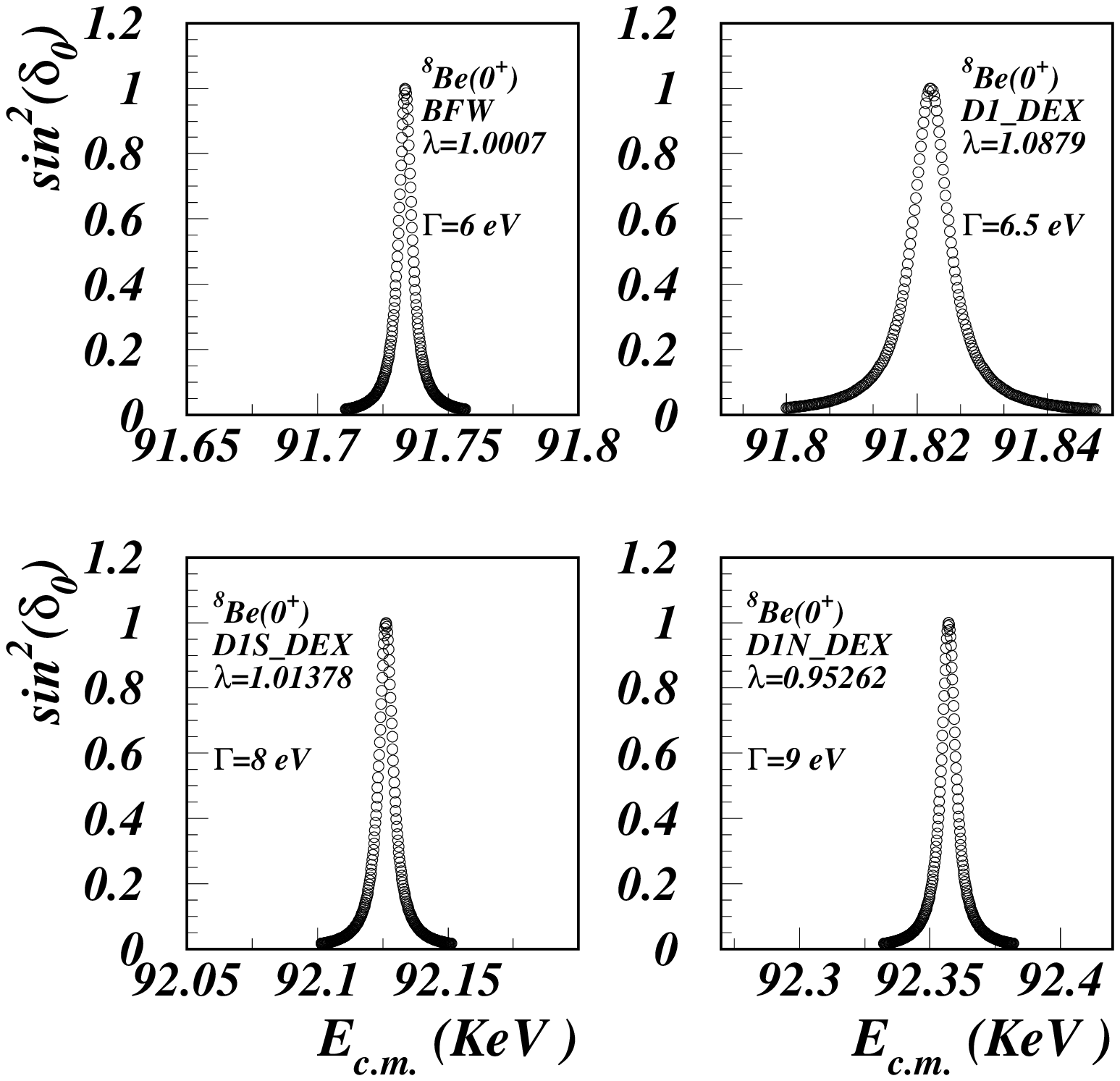}
\label{figpot5}
\caption{The S-state resonance in $^8$Be calculated with the bare folding
potentials. The BFW results are shown for comparison.}
\end{figure}

% insert here figs figpot6, figpot7

\begin{figure}[!ht]
\centering
\includegraphics[width=0.65\textwidth]{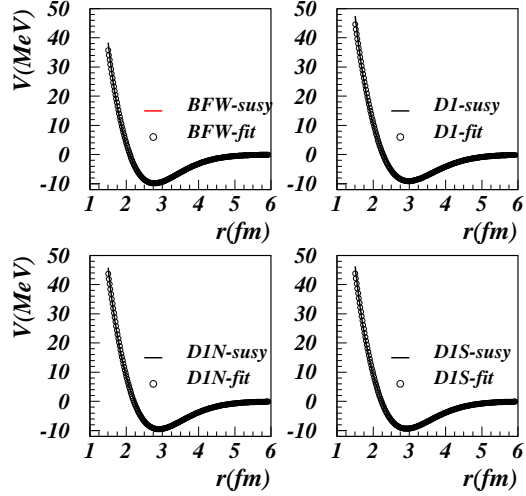}
\label{figpot6}
\caption{Gaussian expansion of the SUSY potentials. The fit was performed in
a restricted radial range $r\sim 1.5-10 fm$}
\end{figure}

\begin{figure}[!ht]
\centering
\includegraphics[width=0.65\textwidth]{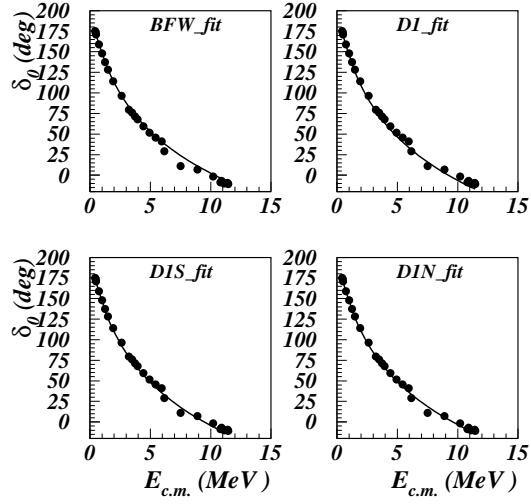}
\label{figpot7}
\caption{ The S-state phase shift calculated with fitted SUSY potentials.}
\end{figure}

\begin{figure}[!ht]
\centering
\includegraphics[width=0.65\textwidth]{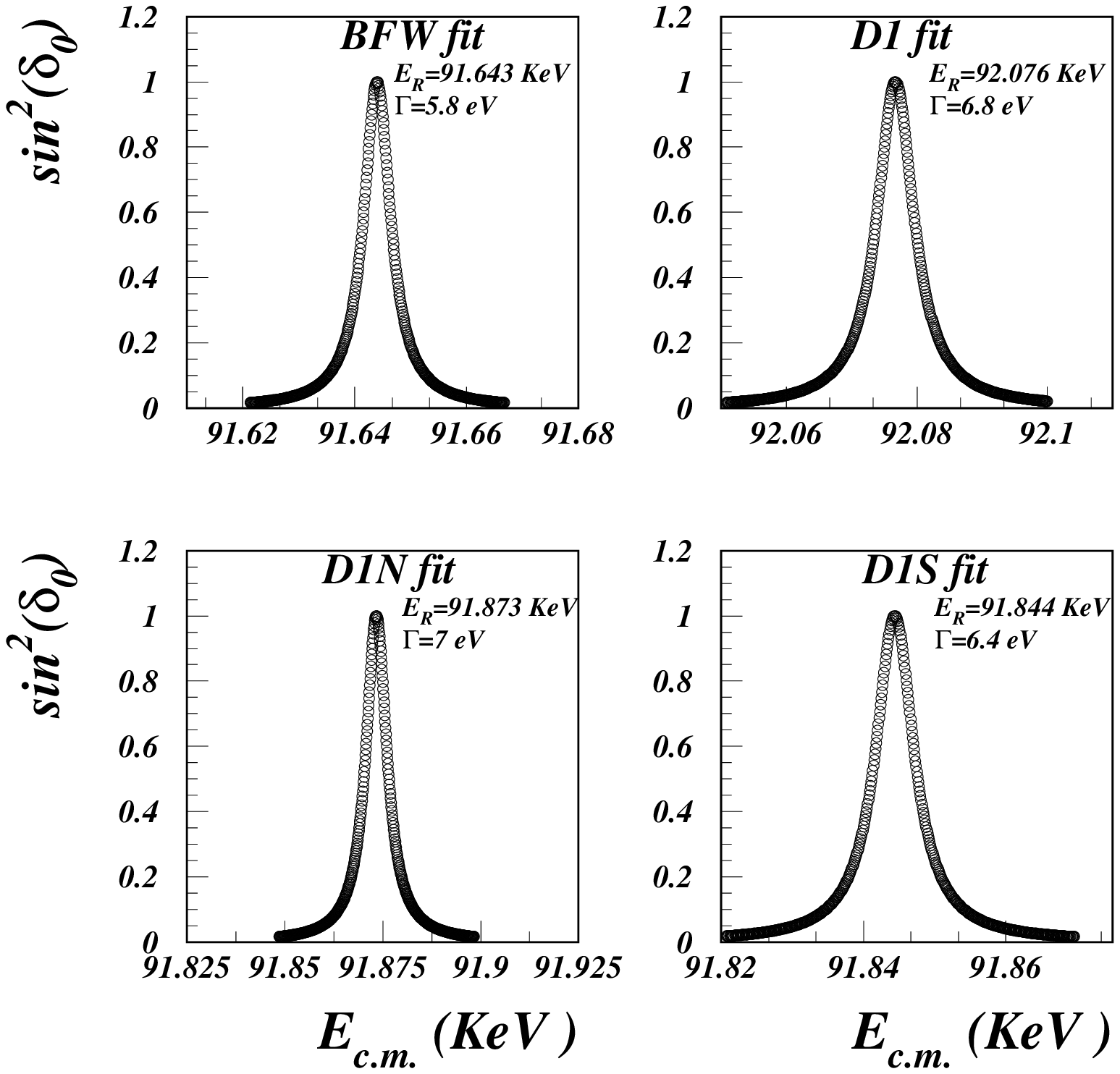}
\caption{The S-state resonance in $^8$Be calculated with fitted SUSY
potentials. Resonance parameters lie in the experimental range \cite{benn} for
all interactions.}
\label{figpot9}
\end{figure}

%\section{Gaussian expansion of the SUSY potentials}
In order to facilitate the calculation for $\alp$-matter we expand the SUSY
potentials in Gaussian form factors, similar to the Ali-Bodmer interaction,

\begin{equation}
V_{fit}(r)=V_r e^{-(\mu_r r)^2}-V_a e^{-(\mu_a r)^2}
\end{equation}
with $V_r,\mu_r,V_a,\mu_a$ fitting parameters. Since it is impossible to obtain
meaningful parameters in the whole radial range, we restrict the fit in the relevant
$r=(1.5-10) $ fm. The result is given in the  Table \ref{susy}. We obtain almost
perfect fits, Fig (6), but comparison with experimental data
require to repeat the renormalization procedure described above. The correction is of the order of $1\%$ in all cases.

\begin{table}[!ht]
\begin{center}
\caption{Parameters for the fitted SUSY potentials. The parameter $\lambda$ is a  renormalization constant which gives the best fit for the experimental S-state phase shift and the $0^+$ resonance in $^8$Be.}
\label{susy}
\begin{tabular}{|c|c|c|c|c|c|}

\hline
Int  &$V_r$(MeV)     &$\mu_r$(fm$^{-1}$)&$V_a$(MeV)                   &$\mu_a$(fm$^{-1}$)&$\lambda$\\\hline
BFW  & 254.8000031   &    0.6470000     &  101.9716263                &   0.4600000      &0.9920\\
 D1  & 255.8999939   &      0.6049346   &  103.6447830                &   0.4370000      &0.9891\\
 D1N & 265.0000000   &      0.6266215   &  102.5655823                &   0.4459522      &0.9873\\
D1S  & 262.0000000   &      0.6194427   &  103.4447250                &   0.4437624      &0.9906\\\hline
\end{tabular}\end{center}\end{table}

\FloatBarrier

\section{Concluding remarks}
We have calculated the $\alp-\alp$ interaction potential within the double
folding model using finite range density dependent NN effective interactions.
The knock-on nonlocal kernel corresponding to the finite range components of the
effective interaction is localized within the lowest order of the Perey-Saxon
approximation at zero energy. The resulted folding potentials are deep with an
average strength of $ 78\pm 7$ MeV very close to the value of Schmid and
Wildermuth \cite{wilderm} in their RGM calculation. The $\it{rms}$ radius of
these potentials is somewhat larger than the corresponding value of
the phenomenological BFW potential (see Fig. 1). Our deep folding potentials
reproduce quite well the experimental values of the S-state phase
shift and the properties of the first $0^+$ resonance in $^8$Be. The maximum
deviation from unity of the usual renormalization factor $\lambda$ is
$9\%$.

 Successive supersymmetric transformations which preserve the continuous spectrum
are used to remove the redundant 0S and 1S states
in order to obtain physically relevant potentials. The phase shift and the properties
of the $0^+$ resonance are calculated with the variable
phase equation of Calogero with proper boundary condition for singular potentials.
A Gaussian expansion of the resulted SUSY potentials shows a well known molecular
pocket with an almost unique long range attractive component with
$\mu_a=0.442\pm 0.005$ fm$^{-1}$. The potential minimum is located at about r=3 fm, which corresponds to a touching configuration and therefore implies a very small overlap of the single particle densities.

We believe that our potentials are physically meaningful in the energy range $E_{lab}=0-5$ MeV.
Beyond this range high $\ell$-order phase shift starts to have significant values.

\section{Acknoledgements}

This work was supported by UEFISCDI-ROMANIA under program PN-II
contract No. 55/2011 and by French-Romanian collaboration IN2P3/IFIN-HH. M. L.
thanks to the staff of DFT/IFIN-HH for the kind hospitality during the
preparation of this work.

\begin{center}

\end{center}

%%%%%%%%%%%%%%%%%%%%%%%%%%%%
\end{document}